\documentclass[twocolumn]{openjournal}
\usepackage[dvipsnames]{xcolor}
\usepackage{aas_macros} 
\usepackage{xr}
\usepackage{amssymb}
\usepackage{amsmath}
\usepackage{graphicx}
\usepackage{hyperref}
\hypersetup{colorlinks=true,linkcolor=blue,citecolor=blue,filecolor=blue,urlcolor=blue}
\usepackage[caption=false]{subfig}
\usepackage{ulem} 
\usepackage{enumerate}
\usepackage{enumitem}
\usepackage{comment}
\pdfoutput=1

\newcommand{\msolar} {$\rm{M_{\odot}}$}
\newcommand{\msolarc} {$\rm{M_{\odot}}$}
\newcommand{\molH} {$\rm{H_2}$}
\newcommand{\molHc} {$\rm{H_2}$}

\begin{document}
\title{Predicting the number density of heavy seed massive black holes due to an intense Lyman-Werner field\vspace{-4em}}
\author{Hannah O'Brennan$^{1,*}$}
\author{John A. Regan$^{1}$}
\author{John Brennan$^1$}
\author{Joe McCaffrey$^{1}$}
\author{John H. Wise$^2$}
\author{Eli Visbal$^{3}$}
\author{Alessandro Trinca$^{4,5,6}$}
\author{Michael L. Norman$^7$}
\thanks{$^*$E-mail:hannah.obrennan.2021@mumail.ie}
\affiliation{$^1$Centre for Astrophysics and Space Sciences Maynooth, Department of Physics, Maynooth University, Maynooth, Ireland}
\affiliation{$^2$Center for Relativistic Astrophysics, Georgia Institute of Technology, 837 State Street, Atlanta, GA 30332, USA}
\affiliation{$^3$Department of Physics and Astronomy and Ritter Astrophysical Research Center, University of Toledo, 2801 W. Bancroft Street, Toledo, OH 43606, USA}
\affiliation{$^4$Como Lake Center for Astrophysics, DiSAT, Università degli Studi dell’Insubria, via Valleggio 11, I-22100, Como, Italy}
\affiliation{$^5$INAF/Osservatorio Astronomico di Roma, Via Frascati 33, 00040 Monte Porzio Catone, Italy}
\affiliation{$^6$INFN, Sezione Roma1, Dipartimento di Fisica, “Sapienza” Universit`a di Roma, Piazzale Aldo Moro 2, 00185, Roma, Italy}
\affiliation{$^7$Center for Astrophysics and Space Sciences, University of California, San Diego, 9500 Gilman Dr, La Jolla, CA 92093, USA}

\begin{abstract}
\noindent The recent detections of a large number of candidate active galactic nuclei at high redshift (i.e. $z \gtrsim 4$) has increased speculation that heavy seed massive black hole formation may be a required pathway. Here we re-implement the so-called Lyman-Werner (LW) channel model of \cite{Dijkstra_2014} to calculate the expected number density of massive black holes formed through this channel. We further enhance this model by extracting information relevant to the model from the \texttt{Renaissance} simulation suite. \texttt{Renaissance} is a high-resolution suite of simulations ideally positioned to probe the high-$z$ Universe. Finally, we compare the LW-only channel against other models in the literature. We find that the LW-only channel results in a peak number density of massive black holes of approximately $\rm{10^{-4} \ cMpc^{-3}}$ at $z \sim 10$. Given the growth requirements and the duty cycle of active galactic nuclei, this means that the LW-only is likely incompatible with recent JWST measurements and can, at most, be responsible for only a small subset of high-$z$ active galactic nuclei. Other models from the literature (e.g. rapid assembly; relative velocities between baryons and dark matter) seem therefore better positioned, at present, to explain the high frequency of massive black holes at high $z$. 
\end{abstract}

\section{Introduction} \label{Sec:Introduction}
\noindent Observations of high-$z$ (i.e. $z \gtrsim 6$) quasars, with masses in excess of $10^8$ \msolarc, indicate that the number density of such massive objects is approximately 1 per cubic gigaparsec \citep{Banados_2018, Inayoshi_2020}. The origin of these high-$z$ quasars poses a major challenge to our understanding of compact object formation and evolution in the early Universe. 
However, these extreme mass objects represent only the tip of the iceberg with smaller mass black holes likely being very much more abundant. Prior to the launch and subsequent observations of the James Webb Space Telescope (JWST), the number densities of this smaller mass population at high redshift could only be inferred from existing relations \citep{Mortlock_2011, Banados_2018, Matsuoka_2019, Venemans_2020, Wang_2021, Yang_2021, Izumi_2021, Andika_2022}. 

\indent Recent observations of massive black holes (MBHs) with masses between $10^6$ and $10^8$ M$_{\odot}$ in the early Universe ($z \gtrsim 4$) have however started to indicate that the population of MBHs is potentially relatively high and certainly at the upper end of what theoretical models had previously suggested \cite[e.g.][]{Habouzit_2016, Greene_2020, Kokorev_2023, Larson_2023, Maiolino_2023, Greene_2024, Harikane_2023}. 
Significant uncertainties in determining the nature of a large population of so-called Little Red Dot (LRD) galaxies mean that it is too early, as of yet, to get clarity on the actual MBH population at high redshift \citep{Lambrides_2024, Li_2024, Ma_2024}. Nonetheless, conservative estimates of the MBH population based on observations of LRDs indicate MBH number densities in excess of $10^{-4}$ cMpc$^{-3}$ \cite[e.g.][]{Perez-Gonzalez_2024} (as opposed to a number density of approximately $10^{-9}$ cMpc$^{-3}$ for the MBH with masses in excess of $10^8$ \msolarc)\footnote{We use the letter 'c' to denote comoving here and so cMpc$^{-3}$ refers to per comoving cubic megaparsec.}.

\indent The question then arises as to what is the formation process that drives the existence of this large population of MBHs? In this work we focus on the so-called heavy seed pathway, where the initial black hole has a mass in excess of $10^3$ M$_{\odot}$ \citep{Regan_2024}. While it is possible that the entire population of MBHs originates from light seeds with initial masses of less than 1000 \msolarc, we do not investigate that scenario here and instead direct interested readers to papers which investigate that channel \citep[e.g.][]{Rees_2001,Alvarez_2009,Madau_2014,Lupi_2016,Smith_2018, Shi_2024, Mehta_2024}. 

\indent For forming heavy seed MBHs, three mainstream (astrophysical) mechanisms have come to the fore which offer possible pathways to achieving the masses of this population within the required time-frame. All of the mechanisms we will discuss, except where we explicitly note, predict the formation of a super-massive star (SMS) as an intermediate stage and subsequently a transition into a MBH. This is often dubbed ``Direct Collapse Black Hole (DCBH)" formation in the literature which is somewhat incorrect given the intermediate stage of stellar evolution which can, in theory, continue for $\geq$ 1 Myr. In fact, a more correct use of the DCBH terminology relates to the concept of the so-called ``Dark Collapse" recently introduced into the literature by \cite{Zwick_2023} and \cite{Mayer_2024}. We will therefore refrain from using the DCBH term here since we focus on the formation of massive stellar objects as precursors to MBH formation. 

\indent The first mechanism theorised to generate MBH seeds is through baryonic streaming velocities. Relative velocity differences between baryons and dark matter will arise following recombination \citep{Tseliakhovich_2010}. While the mean offset will be zero there will, nonetheless, be regions of the Universe where variations from the mean will exist. It is within these regions that the relative velocities can impact the early formation of structure. In particular, streaming motions can act to suppress star formation in the lowest mass halos, pushing the onset of star formation to higher mass halos and perhaps all the way up to the atomic-cooling limit \citep{Naoz_2012, Naoz_2013, Tanaka_2014,Latif_2014c, Hirano_2017, Schauer_2017}. In regions impacted by streaming velocities, the additional velocities of the baryons with respect to the dark matter means that baryons take additional time to settle in the halo centres, thus allowing the halo to grow in mass. Whether realistic streaming motions can truly allow halos to grow to the atomic-cooling limit without triggering star formation is unclear but it may be that the combination of streaming motions with a sufficiently intense Lyman-Werner (LW) flux (described below) may allow for this to occur \citep{Kulkarni_2021, Schauer_2021}. 

\indent The second mechanism is through the rapid growth of structure itself, be it the early rapid assembly of galaxies \citep{Yoshida_2003a} or MBH formation triggered through mergers \citep{Mayer_2010, Mayer_2014, Zwick_2023, Regan_2023}. In the case of early rapid assembly, halos below the atomic-cooling threshold experience rapid minor mergers and hence rapid growth. If the growth is particularly rapid then the heating caused by the dark matter inflows can offset the ability of H$_{2}$ to cool the gas effectively \citep{Fernandez_2014, Wise_2019, Lupi_2021}. In this scenario normal metal-free (Population III; Pop III) star formation is delayed because the gas is unable to cool. Instead the halo continues to accumulate matter without forming stars. 
If the heating (i.e. rapid growth) is maintained then the halo can grow to the atomic-cooling limit where cooling by neutral hydrogen is triggered and star formation occurs regardless. However, having cooling commence at this mass scale offers the opportunity for SMS formation to occur as the gravitational potential is now sufficiently deep to enable this mechanism \citep[e.g][]{Regan_2020b, Latif_2022, Regan_2023}. 

\indent In addition to this pathway, the related pathway of major mergers can drive huge gas inflows directly into the centres of merging halos. The large gas inflows can in some cases initiate a dark collapse through the formation of a super-massive disk \citep{Zwick_2023} and ultimately the direct formation of a MBH \citep{Mayer_2010, Mayer_2024}. While the investigations of \cite{Mayer_2024} focused on very massive objects and the formation in particular of high-$z$ quasars, it is likely that this mechanism also acts on smaller mass scales, leading to the formation of a population of high-$z$ MBHs with masses closer to the expected MBH seed masses \citep{Regan_2023}.

\indent The final mechanism and the one we will focus on in this paper is driven by local sources of LW radiation. LW radiation is emitted by stars and is composed of radiation below the hydrogen ionisation edge at photon energies between 11.2 and 13.6 eV. In order to form a Pop III star, the gas within a mini-halo\footnote{We use the term mini-halo to refer to halos above $\rm{M_{halo}} = 10^5$ M$_{\odot}$ and below the atomic-cooling threshold.} must cool down to T$_{\text{gas}} \approx$ 200 K, allowing the gas to achieve the required densities and pressures to ignite star formation. Sufficiently intense LW radiation works against this process by dissociating H$_{2}$, thus removing (or at least suppressing) a critical coolant required for star formation to take place. In the primordial Universe, the formation of H$_{2}$ takes place via two possible routes \citep{Galli_1998}. The first is through the radiative association of $\rm{H}$ and $\rm{H^+}$ below: 
\begin{equation}
\rm{H + H^+ \rightarrow H_2 + \gamma}.
\end{equation}
This reaction is only important at very high redshift ($z \gtrsim 365$) when T$_{\text{CMB}} > 10^3$ K and hence is less relevant for galaxy formation. The more relevant reaction is the associative detachment reaction of $\rm{H}$ and $\rm{H^-}$ which gives
\begin{equation}
\rm{H + H^- \rightarrow H_2 + e^-}.
\end{equation}
This is the critical reaction of H$_{2}$ formation in the high-$z$ Universe relevant for galaxy formation and depends sensitively on the $\rm{H^-}$ abundances. In opposition to these formation pathways is LW radiation which can dissociate H$_{2}$. Radiation in the LW band excites electrons in the H$_{2}$ molecule, breaking the molecule down into its constituent atoms as such:
\begin{equation}
\rm{H_2 + \gamma_{LW} \rightarrow H_2^* \rightarrow H + H},
\end{equation}
where initially the LW photons excite electrons in the H$_{2}$ molecule into an excited state from where they can decay with some probability into two $\rm{H}$ atoms. Crucially by removing the (inefficient) coolant \molH, the gas cannot cool and hence star formation is suppressed until the halo reaches the atomic-cooling limit, allowing the gas to cool and condense via neutral hydrogen transitions. 

\indent The intensity of LW radiation required to achieve full star formation suppression in mini-halos has been well-studied over the past decade. The consensus is that, while there is some spectral dependence on the critical flux of LW radiation required, values of at least 300 $\rm{J_{21}}$ are required for a $\rm{T} \gtrsim 10^4 $ K spectrum \citep[e.g.][]{Latif_2014, Regan_2017} while values of as high as $\rm{1000 \ J_{21}}$ may be required for a spectrum composed of Pop III stars only (i.e. $\rm{T \sim 10^5}$ K spectrum) \citep{Agarwal_2015, Agarwal_2016} (where $\text{J}_{21} = 10^{-21} \, \text{erg} \, \text{s}^{-1} \, \text{Hz}^{-1} \, \text{sr}^{-1} \, \text{cm}^{-2}$). The value of the LW flux required is then known as the critical flux, $J_{\text{crit}}$. The question then becomes under what circumstances can a LW flux greater than $J_{\text{crit}}$ be achieved in practice? The goal of this paper is to re-examine the methodology introduced by \cite{Dijkstra_2008, Dijkstra_2014} (hereafter D08 \& D14) while also augmenting the models with data from the \texttt{Renaissance} \citep{Chen_2014, OShea_2015, Xu_2016b} simulation suite. Finally, we also compare the LW-only channel against more recent estimates of MBH heavy seed number densities through other channels \citep[e.g.][]{Trinca_2022, McCaffrey_2024}. Using these combined results, we put tight limits on the number density of MBH seeds formed via LW feedback only and critically examine how viable the LW-only pathway is in the face of recent JWST observations. 
The structure of the paper is as follows: In \S \ref{Sec:Methodology}, we outline the methodology including the model from D14 and the relevant output post-processed from \texttt{Renaissance}. In \S \ref{Sec:Results}, we deliver the results of our analysis, showing an updated heavy seed number density plot. In \S \ref{Sec:Discussion}, we summarize and discuss our results in light of recent JWST observations and other models in the literature.

\section{Methodology} \label{Sec:Methodology}
\noindent In this section, we detail our application of the heavy seed formation model used by D08 and D14. In \S \ref{subsec:heavy_seed_formation}, we describe the conditions a halo must meet to potentially host a heavy seed. In \S \ref{subsec:renaissance}, we introduce the \texttt{Renaissance} simulations and how we post-process the data to supplement the fiducial (analytic-only) model. In \S \ref{subsec:hmf}, we define the halo mass function and how we use it to compute the halo number density. In \S \ref{subsec:genetic_metal_pollution}, we discuss the impact of metal pollution, both from merger history and from neighbouring galaxy outflows. Finally, in \S \ref{subsec:LW_flux}, we show how the supercritical LW flux probability is computed, both in the fiducial model and one informed by \texttt{Renaissance} data.

\subsection{Formation of Heavy Seeds} \label{subsec:heavy_seed_formation}
\noindent Here we refer to heavy seeds as MBH seeds that have initial masses $\geq$ $10^3$ \msolarc, as opposed to light seeds with initial masses $<$ $10^3$ \msolarc. We find the heavy seed number density $n_{\text{heavy seeds}}(z)$ (units: cMpc$^{-3}$) as a function of redshift $z$ ($10 \leq z \leq 30$) by computing the number density of dark matter halos which meet the criteria to host heavy seeds (assuming one heavy seed forms per host halo). For the purpose of this study, the dominant criteria is the impact of a LW radiation field as discussed in \S \ref{Sec:Introduction} and in detail below. 

\begin{figure}
    \centering
    \includegraphics[scale=0.5]{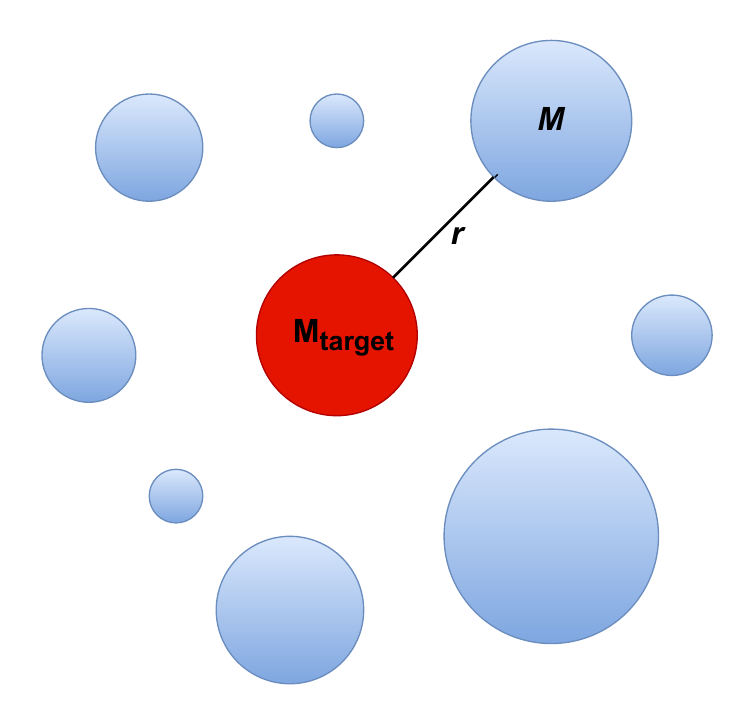}
    \caption{Target halo surrounded by neighbouring halos of various masses and physical separations. Here $M$ refers to the mass of a given neighbouring halo, $r$ is the physical separation between the neighbouring halo and the (central) target halo. The central target halo is later defined as having a mass M$_{\text{target}} = \text{M}_{\text{min}}(z)$ which
    is exactly equivalent to the virial mass of a halo at redshift $z$ with T$_{\text{vir}} = 10^4$ K.}
    \label{fig:Target_halo_with_neighbours}
\end{figure}

Consider a potential host halo of mass M$_{\text{target}}$ at redshift $z$ (see Figure \ref{fig:Target_halo_with_neighbours}). It is surrounded by neighbouring dark matter halos of various masses, $M$, and physical separations, $r$. At this high-redshift range, the comoving distance from an observer on Earth to such a host halo is $\geq$ 9 cGpc (at $z=10.0$, the proper distance would be $\gtrsim$ 800 Mpc). We are considering physical separations between the host halo and its neighbours of $\lesssim$ 10 Mpc. Since these separations are negligible compared to the proper distances from Earth, we approximate the neighbouring halos as having the same redshift as the potential host.

To form light seeds within a halo, baryonic gas would cool and fragment to form stars which would later collapse into stellar black holes\footnote{At time of writing, it is still not clear whether there is a mass gap between the most massive neutron stars ($\approx 2$ M$_{\odot}$, Tolman-Oppenheimer-Volkoff limit \citep{Heger_2003}) and a minimum stellar black hole mass ($\approx 5$ M$_{\odot}$) \citep{Farr_2011, Kreidberg_2012, Abbott_2019}.}. Thus to reduce fragmentation and the formation of light seeds (forming heavy seeds instead), the temperature of the gas within the target halo must remain high ($\rm{T_{gas}} \gtrsim 8000$ K). 

This can be achieved if the target halo meets the following criteria:
\begin{enumerate}[itemsep=0.5pt,parsep=1pt]
    \item it is massive enough for the gas to cool via atomic hydrogen only;
    \item it is chemically pristine i.e. free from metal enrichment;
    \item it receives sufficient LW radiation from within its neighbours to suppress H$_{2}$ cooling.
\end{enumerate}

The number density of dark matter halos (units: cMpc$^{-3}$) meeting these criteria (and thus the heavy seed number density) is described as:
\begin{align}\label{eqn:n_IMBH_den}
    n_{\text{heavy seeds}}(z) = &\int_{\text{M}_{\text{min}}(z)}^{\infty} d\text{M}_{\text{target}} \frac{dn_{\text{SMT}}}{dM}(z, \text{M}_{\text{target}}) \\
                             &\times P_{\text{pristine}}(z) \, P_{\text{LW}}(z, \text{M}_{\text{target}}). \nonumber
\end{align}
We now briefly define each term in the above integral. We integrate over M$_{\text{target}}$, the mass of a halo that is a potential heavy seed (or equivalently MBH) formation site. Here M$_\text{min}(z)$ is the minimum mass where the virial theorem is satisfied with T$_{\text{vir}} = 10^{4}$ K \citep{Barkana_2001} i.e. the atomic-cooling limit and is given as:  
\begin{equation}
    \text{M}_{\text{min}}(z) = 4 \times 10^{7} \left(\frac{1+z}{11}\right)^{-3/2} \, \text{M}_{\odot}.
\end{equation}
We integrate the halo mass function $\frac{dn_{\text{SMT}}}{dM}(z, \text{M}_{\text{target}})$ (units: $\text{M}_{\odot}^{-1}$ cMpc$^{-3}$) to find the number density of halos greater than or equal to the atomic-cooling limit. The halo mass function gives the number of halos per comoving volume per unit mass. It was originally derived analytically in a seminal paper by \cite{PS_1974}. In this work we use the semi-analytic form from \cite{SMT_2001} (SMT) which assumes ellipsoidal halo collapse (as opposed to the idealised spherical collapse assumed by \cite{PS_1974}). We implement it using the Python package \texttt{hmf}, developed by \cite{Murray_2013}.

The quantity $P_{\text{pristine}}(z)$ refers to the probability of a target halo being pristine i.e. free from metal pollution. In \S \ref{subsec:genetic_metal_pollution}, we discuss both the model used by D14 \citep{Trenti_2007,Trenti_2009} and one derived from \texttt{Renaissance} data and refer to the derived probabilities as $P_{\text{pristine, fid.}}(z)$ and $P_{\text{pristine, Ren.}}(z)$ respectively. $P_{\text{pristine, fid.}}(z)$ accounts only for metal pollution inherited via merging episodes. $P_{\text{pristine, Ren.}}(z)$ accounts for both inherited metal pollution and metal enrichment via supernovae outflows from neighbouring halos and does not distinguish between the two.

Finally, a given target halo receives LW radiation from its neighbours (which dissociates $\rm{H_2}$) but also metal outflows from supernovae (which act as a coolant). If the target halo receives a total LW flux exceeding a threshold flux ($J > J_{\text{crit}}$), it is a candidate for heavy seed formation due to the sufficient dissociation of \molHc. The quantity $P_{\text{LW, fid.}}(z, \text{M}_{\text{target}})$ refers to the probability of a target halo at redshift $z$ and of mass M$_{\text{target}}$ receiving supercritical LW radiation while also avoiding metal outflows from its neighbouring halos in the fiducial model. The analogous quantity $P_{\text{LW, Ren.}}(z, \text{M}_{\text{target}})$ refers only to supercritical LW radiation as metal outflows are accounted for by $P_{\text{pristine, Ren.}}(z)$ (see \S \ref{subsec:genetic_metal_pollution} for more detail). In this model, $J_{\text{crit}}$ is a parameter and we evaluate the supercritical probability, and subsequently the heavy seed number density, at $J_{\text{crit}} = $ 300 and 1000 $\text{J}_{21}$. This probability is highly dependent on the form of the mean LW luminosity density $\langle L_{\text{LW}}(z, M)\rangle$. We investigate both the form described by D14 and one derived from \texttt{Renaissance} data, referred to as $\langle L_{\text{LW, fid.}}(z, M)\rangle$ and $\langle L_{\text{LW, Ren.}}(z, M)\rangle$ respectively.

We describe each of the terms in Eq. \ref{eqn:n_IMBH_den} in greater detail in \S \ref{subsec:hmf}, \S \ref{subsec:genetic_metal_pollution} and \S \ref{subsec:LW_flux} below. Our cosmology matches the cosmology used by D14: $n_{\text{s}} = 0.9624$, $\text{h} = 0.6711$, $\Omega_{\text{m},0} = 0.3175$, $\Omega_{\text{b}, 0}\text{h}^{2} = 0.022068$, $\Omega_{\Lambda} = 0.6825$ and $\sigma_{8} = 0.8344$. The impact of a specific cosmology on our model is however likely to be minimal. Note that we use units with factors of h when performing calculations. This is to be consistent with units used in \texttt{hmf}\footnote{hmf: https://pypi.org/project/hmf/1.6.2/} and \texttt{halomod}\footnote{halomod: https://github.com/halomod/halomod/}, two Python packages that we use to compute the halo mass function, power spectrum and halo bias factor \citep{Murray_2013, Murray_2021}. Our final results in Figure \ref{fig:Number_density_plot} have been converted to units without factors of h.

\subsection{\texttt{Renaissance} Simulations}\label{subsec:renaissance}
\noindent The \texttt{Renaissance} simulations \citep{Xu_2013, Xu_2014, Chen_2014, OShea_2015, Smith_2018,Wise_2019} were run using the massively parallel adaptive mesh refinement \texttt{Enzo} code \citep{Enzo_2014, Enzo_2019}. We briefly describe the \texttt{Renaissance} suite here, but refer the interested reader to the previous papers for a more complete discussion. The \texttt{Renaissance} simulation suite is composed of three zoom-in regions extracted from a parent volume of (40 cMpc)$^{3}$. The three separate zoom-in regions were named the \textit{Rarepeak} (RP) region, the \textit{Normal} region and the \textit{Void} region.  Each volume was smoothed on a scale of 5 cMpc, with the RP region corresponding to a mean overdensity of $\langle \delta \rangle \equiv \frac{\langle \rho \rangle}{\Omega_{\text{m}} \rho_c} -1 \sim 0.68$ and the \textit{Normal} region corresponding to $\langle \delta \rangle \equiv \frac{\langle \rho \rangle}{\Omega_{\text{m}} \rho_c} -1 \sim 0.09$. The RP and \textit{Normal} subregions have volumes of 133.6 and 220.5 cMpc$^3$ respectively. They were resimulated with an effective initial resolution of 4096$^3$ grid cells and particles within the central, most refined regions, and have a particle mass resolution of $2.9 \times 10^4$ \msolarc. In addition to the initial nesting procedure, which allows the RP and \textit{Normal} regions to be selected, up to 8 further levels of refinement are allowed, giving a total maximum spatial resolution of $\approx$ 19 cpc. For this study we focus on the \textit{Normal} region only since it is a representative volume of mean cosmic density. This is potentially quite conservative since it is less likely to find high values of $P_{\text{LW}}(z, J_{\text{crit}})$ within \textit{Normal} regions. We make this choice since the alternatives (RP and \textit{Void}) would be difficult to relate to observations.

From \texttt{Renaissance} we can compare some of the key characteristics of the D14 model against simulation datasets in order to test the veracity of some of the key underlying assumptions of the model. Considering again Eq. \ref{eqn:n_IMBH_den}, we see that calculating the number density depends on the halo mass function, the probability of a halo receiving supercritical flux and the probability of a halo remaining pristine. From \texttt{Renaissance} the key quantities that can be extracted are the mean LW luminosity densities $\langle L_{\text{LW, Ren.}}(z, M)\rangle$ as a function of redshift and neighbouring halo mass and the pristine fraction of halos $P_{\text{LW, Ren.}}(z, \text{M}_{\text{target}})$ as a function of redshift and target halo mass. 

Note that while previous works (see e.g. \cite{Smith_2018, McCaffrey_2024}) showed that \texttt{Renaissance} post-processed black holes struggled to grow to heavy seed scales i.e. $M_{\text{BH}} \sim 10^5$ M$_{\odot}$, we are only interested in halos themselves as hosts whose environment fits the criteria to later form heavy-seed black holes. We do not investigate the later growth and dynamics of these black holes in this study.

However, we begin our analysis by examining the halo mass function, which is the main driver of the number of heavy seeds. We then move onto the pristine fraction (where \texttt{Renaissance} will play a role) and the probability of a halo receiving a super-critical flux (where again \texttt{Renaissance} will play a role). 

\subsection{Halo Mass Function}\label{subsec:hmf}
\begin{figure}
    \centering
    \includegraphics[scale=0.45]{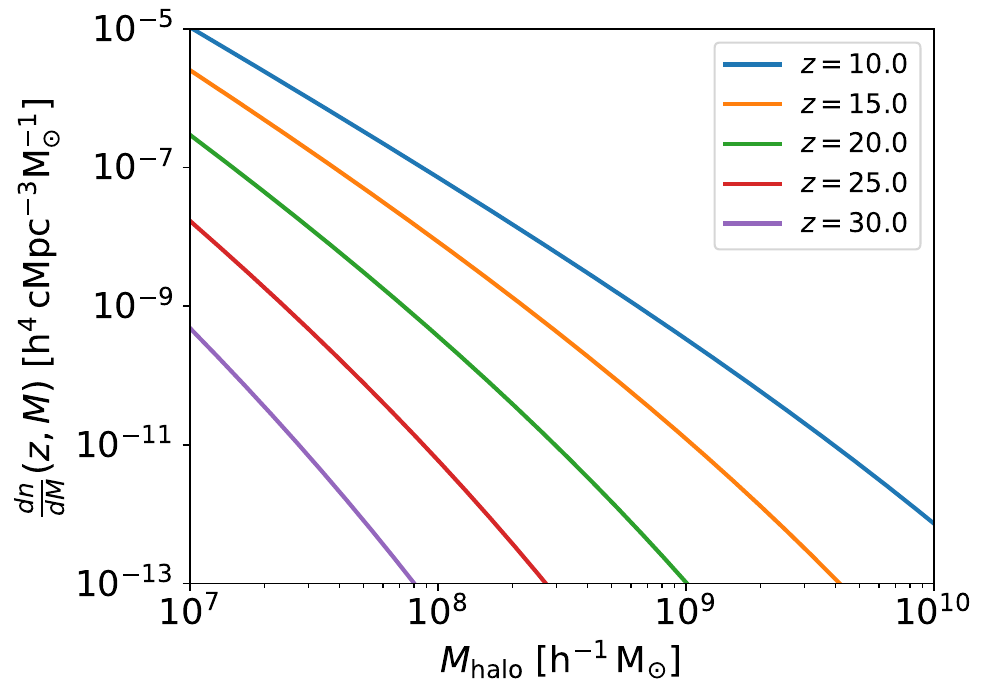}
    \caption{Halo mass function vs. halo mass at a number of redshifts. The lines shown here are generated using the \texttt{hmf} package developed by \cite{Murray_2013}. We use the SMT halo mass function and the modified Planck13 cosmology (used also by D14). The rarity of halos in a given mass range increases with increasing redshift.}
    \label{fig:hmf_plot}
\end{figure}

\noindent In Figure \ref{fig:hmf_plot} we show how the halo mass function varies with halo mass ($10^{7} \, \text{h}^{-1} \, \text{M}_{\odot} \leq \text{M}_{\text{target}} \leq 10^{10} \, \text{h}^{-1} \, \text{M}_{\odot}$) at a number of fixed redshifts. More massive halos become significantly rarer at all redshifts, with the higher redshifts showing increased rarity of high mass halos as expected in a $\Lambda$CDM cosmology. The halo mass function is relatively well calibrated against N-body simulations at lower redshifts but can differ from N-body runs at high-$z$ by up to an order of magnitude \citep[for a discussion on this point see][]{Yung_2023,O_Brennan_2024}. However, for the halo masses which we are concerned with here (i.e. close to the atomic-cooling limit) the differences are approximately a factor of two and we do not investigate systematic differences due to the halo mass function here.

Since the halo mass function decreases by several orders of magnitude with increasing halo mass, we approximate Eq. \ref{eqn:n_IMBH_den} as:
\begin{align}\label{eqn:n_IMBH_den_approx}
    n_{\text{heavy seeds}}(z) &\approx \int_{\text{M}_{\text{min}}(z)}^{\infty} d\text{M}_{\text{target}} \frac{dn_{\text{SMT}}}{dM}(z, \text{M}_{\text{target}}) \nonumber \\
                              &\times P_{\text{pristine}}(z)\, P_{\text{LW}}(z, \text{M}_{\text{min}}(z)), \nonumber \\
                              & \\
    n_{\text{heavy seeds}}(z) &\approx n_{\text{halo}}(\text{M}_{\text{target}} > \text{M}_{\text{min}}(z)) \nonumber \\
                              &\times P_{\text{pristine}}(z) \times P_{\text{LW}}(z, \text{M}_{\text{min}}(z)), \nonumber
\end{align} i.e. when computing $P_{\text{LW}}$ in \S \ref{subsec:LW_flux}, we approximate all potential formation sites at some redshift $z$ as having a fixed mass of M$_{\text{target}} = \text{M}_{\text{min}}(z)$.

\subsection{Metal Pollution}\label{subsec:genetic_metal_pollution}
\noindent Metal pollution can come in two forms. Firstly it can be inherited ``genetically" through the hierarchical assembly process and secondly it can come from outflows from neighbouring halos. For the fiducial (analytic-only) model these two forms are computed separately, while for the model where we additionally use \texttt{Renaissance} data we simply calculate the halo metallicity (or more precisely the probability of a halo being metal-enriched via mergers or outflows from neighbours).

\subsubsection{Genetic Metal Pollution}
\noindent Starting with the analytic-only model: consider a target halo at redshift $z$ with mass M$_{\text{target}}$. During previous episodes of halo merging, this target halo may have had a progenitor halo that became metal-enriched via supernova outflows from Pop III stars. If so, then the target halo would inherit this metal pollution i.e. genetic metal pollution. The quantity $P_{\text{pristine, fid.}}(z)$ refers to the probability of the target halo avoiding this genetic metal pollution i.e. it does not have a metal-enriched progenitor halo.

\begin{figure}[h]
    \centering
    \includegraphics[scale=0.5]{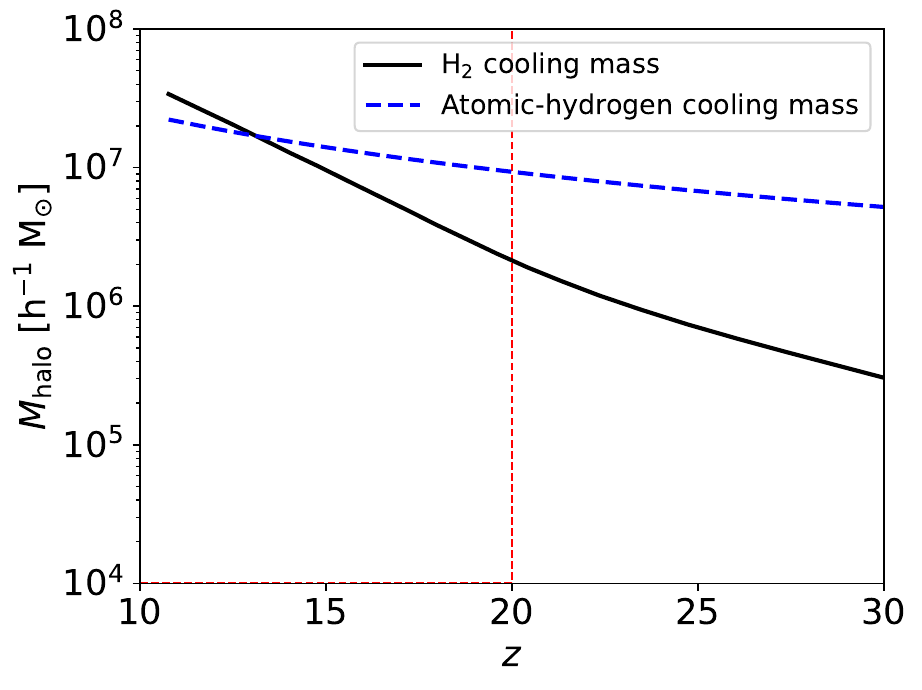}
    \caption{A recreation of Figure 1 (upper right panel) from \cite{Trenti_2009}. In this redshift range, the minimum halo mass for H$_{2}$ cooling to occur (solid black line) sharply increases as $z$ decreases and is greater than the atomic-cooling limit (blue dashed line) at $z \lesssim 13$. The red dashed vertical line marks $z = 20$, the highest redshift considered by D14.}
    \label{fig:TS09_Fig1_UR}
\end{figure}

In our fiducial model, as with D14, we base the form of $P_{\text{pristine, fid.}}(z)$ on the model proposed by \cite{Trenti_2007} (TS07) and \cite{Trenti_2009} (TS09). TS07 investigated the formation of the first generation of Pop III stars within dark matter halos and tracked their descendant halos using a combination of \texttt{Gadget-2} N-body simulations and a Monte Carlo method based on linear theory. Using these simulation results, TS09 found how M$_{\text{H}_2}(z, J_{\text{bg}})$ (units: h$^{-1}$ M$_{\odot}$) varied as a function of redshift $z$ and LW background $J_{\text{bg}}$. M$_{\text{H}_2}$ refers to the minimum mass where H$_{2}$ cooling may occur within a progenitor halo, and later Pop III star formation.\footnote{TS09 found $J_{\text{bg}}(z)$ as a function of $z$, so M$_{\text{H}_2}$ is strictly a function of $z$ only.} Figure \ref{fig:TS09_Fig1_UR} (a recreation of TS09 Figure 1 (upper right panel)) shows how this mass increases by  a factor of $\approx 30$ as redshift decreases from $z \approx 20$ to $z \approx 10$. Conversely, the atomic-cooling limit M$_{\text{min}}(z)$ increases only by a factor of $\approx 4$ in the same redshift range and M$_{\text{min}}(z) < \text{M}_{\text{H}_2}(z, J_{\text{bg}})$ at $z \lesssim 13$. Thus a progenitor halo is more likely to cool via atomic hydrogen than H$_{2}$ at low redshifts. 

\begin{figure}
    \centering
    \includegraphics[scale=0.55]{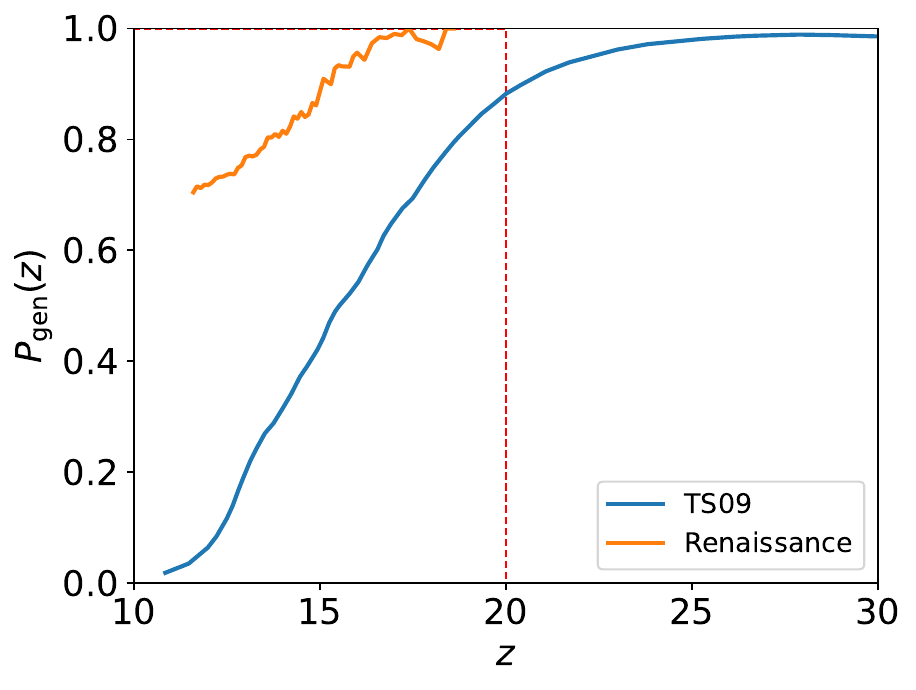}
    \caption{A recreation of Figure 1 (lower right panel) from \cite{Trenti_2009}. This is the probability for genetic metal pollution to occur within a halo with T$_{\text{vir}} = 10^{4}$ K. From $z=20$ to $z=10$, $P_{\text{gen}}(z)$ (solid blue line) sharply decreases i.e. a target halo at low redshift is less likely to inherit metal pollution from a progenitor halo. Overplotted as the solid orange line is the probability of a halo with mass in excess of the atomic-cooling mass being metal polluted taken from the \texttt{Renaissance} simulation suite. This metal pollution probability includes receiving metals from both genetic metal pollution and external metal pollution. The dashed red line marks $z$ = 20.}
    \label{fig:TS09_Fig1_LR}
\end{figure}

\begin{figure}
    \centering
    \includegraphics[scale=0.55]{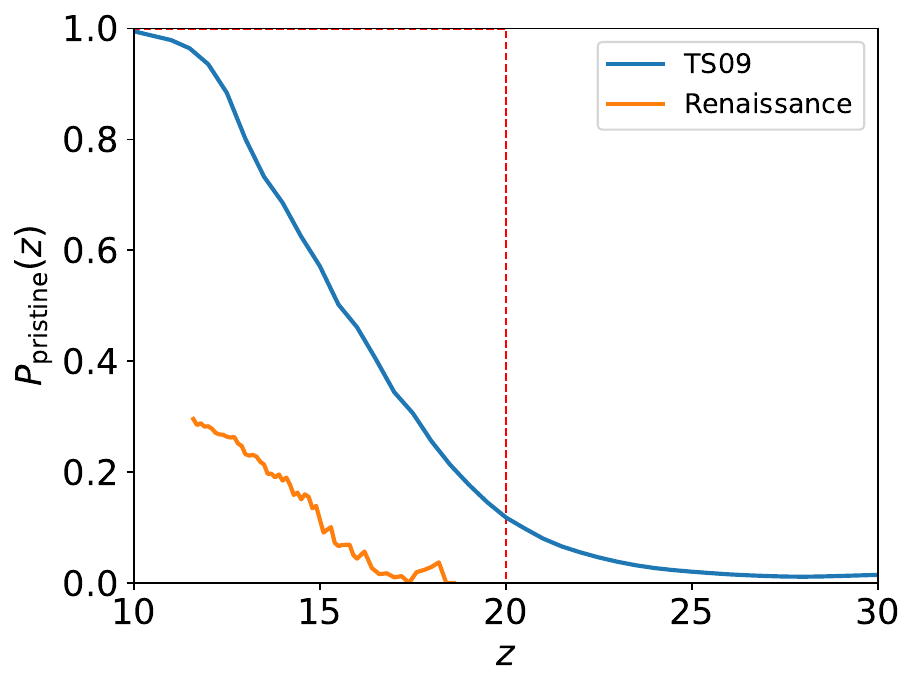}
    \caption{$P_{\text{pristine}}(z) = 1 - P_{\text{gen}}(z)$ as a function of redshift, based on the model used by TS07 and TS09 (solid blue line). From $z=20$ to $z=10$, $P_{\text{pristine, fid.}}(z)$ sharply increases i.e. a target halo at low redshift is more likely to be free of genetic metal pollution. In orange we plot $P_{\text{pristine, Ren.}}(z)$ from the \texttt{Renaissance} simulation suite (which again accounts for both genetic and external metal enrichment). The dashed red line marks $z$ = 20.}
    \label{fig:P_pristine}
\end{figure}

\noindent As a result, TS09 have shown that the probability of Pop III star formation is less likely in a progenitor halo as redshift decreases. Subsequently the probability, $P_{\text{gen}}(z)$, of a target halo inheriting metal pollution decreases as redshift decreases (see our Figure \ref{fig:TS09_Fig1_LR} for a recreation of TS09 Figure 1 (lower right panel)). TS09 found that $P_{\text{gen}}(z=20) \approx 0.9$ and $P_{\text{gen}}(z=10) \approx 0.1$.

We now look at metal enrichment in \texttt{Renaissance}. For the \texttt{Renaissance} halos, we select only halos above the atomic-cooling threshold mass, $\rm{M_{min}}$, and from the \textit{Normal} region only. We show the probability of a halo being metal-enriched in \texttt{Renaissance} in Figure \ref{fig:TS09_Fig1_LR} as the orange line i.e. the fraction of halos found to be metal-enriched at a given redshift. Note that this metal enrichment is from both genetic pollution and external enrichment and hence direct comparisons should be treated with this in mind. We consider such halos to be pristine only if they are metal-free i.e. $Z < 10^{-16}$ Z$_{\odot}$. Halos with greater metallicity are counted as metal-enriched. At high redshifts ($z \gtrsim 18$), the fraction is zero due to the fact that there are no halos with masses in excess of $\rm{M_{min}}$ at that point in the simulation (sub)volumes. As structure formation evolves and halos accrete sufficient mass, then the first atomic halos are predominantly star-forming and metal-enriched and hence the fraction of those halos which are pristine is exceptionally low to start (see Figure \ref{fig:P_pristine}). As more and more halos form then the number (and fraction) of pristine halos increases substantially approaching 30\% by the end of the simulation. No turnover is seen but this is expected to occur as metal diffusion becomes more widespread over cosmic time. Note that the apparent sharp decrease in $P_{\text{pristine}}$ from $z \approx 18$ to 17 in Figure \ref{fig:P_pristine} is likely noise due to small number statistics i.e. there are very few halos in this redshift range that are massive enough to be counted in this analysis.

The \texttt{Renaissance} results are clearly deviant from the analytical models - likely due to the self-consistent treatment of structure formation and metal enrichment available to the hydrodynamic simulations. However, as noted already, we need to be somewhat careful here. Metal enrichment in the \texttt{Renaissance} halos can come from either genetic enrichment as well as external enrichment (via outflows from neighbouring galaxies). While we see a similar trend to the analytic models of TS07 and TS09, we clearly see that the hydrodynamical simulations predict higher values of (genetic and external) metal pollution than the analytic models. Figure \ref{fig:P_pristine} shows the same result, albeit inverted since we show the pristine fraction (which is required by our model). For the analytic model we address the issue of external metal enrichment in \S \ref{subsubsec:metalneighbours} below.

\subsubsection{Metal Pollution from Neighbouring Halos} \label{subsubsec:metalneighbours}
Within a neighbouring halo of mass $M$ (units: h$^{-1}$ M$_{\odot}$), massive stars eject metals as supernovae at the end of their lives. We approximate this phenomenon by assuming a fraction of stars within the halo enter their supernova phase simultaneously and there is a single physical radius of metal pollution $r_{\text{s}}(z, M, t)$ (units: h$^{-1}$ Mpc) growing from the halo centre as $t$ increases. It is given as:

\begin{equation}
    r_{\text{s}}(z, M, t) = \Big(\frac{E_{0}\nu f_{*}\Omega_{\text{b},0}M}{\rho_{\text{gas}}(z)\Omega_{\text{m},0}}\Big)^{1/5}t^{2/5},
\end{equation}
where $E_{0} = 10^{51}$ erg is the supernova explosion energy, $\nu = 0.01/\text{h}$ h $\text{M}_{\odot}^{-1}$ is the number of supernova per unit mass formed and $\rho_{\text{gas}}(z) = \Delta\Omega_{\text{b},0}\rho_{\text{crit}, 0}(1+z)^{3}$ is the density of the ambient gas. We set our over-density parameter to $\Delta=60$. We assume that the beginning of the target halo collapse, the beginning of star formation within the neighbouring halo and the beginning of the neighbouring supernova phase occur simultaneously (at $t=0$). The target halo must avoid metal pollution while $t < t_{\text{ff}}(z)$ where $t_{\text{ff}}(z)$ is the free-fall time of the target halo. In other words for a set redshift $z$ and neighbouring halo mass $M$, the halo separation $r > r_{\text{s}}(z, M, t=t_{\text{ff}}(z))$ since $r_{\text{s}}$ increases monotonically with $t$. When computing the heavy seed number density in the fiducial model, we include a Heaviside function $\Theta[r - r_{\text{s}}(z, M, t_{\text{ff}}(z))]$ to only include metal-free halos described above. This is only necessary for the analytic model. When modelling using the \texttt{Renaissance} data, we do not need to include the Heaviside function as this information is already contained as part of the pristine fraction of halos $P_{\text{pristine, Ren.}}(z)$.

\subsection{Supercritical Flux}\label{subsec:LW_flux}
Here we derive $P_{\text{LW}}(z, \text{M}_{\text{target}})$ i.e. the probability of a target halo receiving supercritical LW radiation. Since we use the approximation described in Eq. \ref{eqn:n_IMBH_den_approx}, this probability is now a function of $z$ only i.e. $P_{\text{LW}}(z, \text{M}_{\text{target}}) = P_{\text{LW}}(z, \text{M}_{\text{min}}(z)) = P_{\text{LW}}(z)$. We compute this analytically by approximating that the LW flux received by the target halo is dominated by a single luminous nearby source. 

We integrate a probability density over LW flux $J$:
\begin{equation}
\begin{aligned}
P_{\text{LW, fid.}}(z) &= \int_{\log_{10}J_{\text{crit}}}^{\infty} d\log_{10} J \, \frac{dP_{\text{fid.}}}{d\log_{10}J}(z, J),\\
P_{\text{LW, Ren.}}(z) &= \int_{\log_{10}J_{\text{crit}}}^{\infty} d\log_{10} J \, \frac{dP_{\text{Ren.}}}{d\log_{10}J}(z, J).
\end{aligned}
\end{equation}
Here $J$ is in units of J$_{21}$ and we integrate over the logarithmic value. The LW flux probability density is found by integrating over all possible masses (units: h$^{-1}$ M$_{\odot}$) and physical separations (units: h$^{-1}$ Mpc) of neighbouring halos:
\begin{align}\label{eqn:dP_dlog10J}
\frac{dP_{\text{fid.}}}{d\log_{10}J}(z, J) &= \int_{M_a}^{M_{b, \text{fid.}}} dM  \int_{r_{\text{min}}(z,M)}^{r_{\text{max}}(z)}  dr \frac{d^{2}P}{dMdr} \nonumber \\
                                                &\times \frac{dP}{d\log_{10}L}(\langle L_{\text{LW, fid.}}\rangle, J, r) \nonumber \\ 
                                                &\times \Theta[r - r_{\text{s}}(z, M, t_{\text{ff}}(z))], \nonumber\\
                                                \\
\frac{dP_{\text{Ren.}}}{d\log_{10}J}(z, J) &= \int_{M_a}^{M_{b, \text{Ren.}}} dM  \int_{r_{\text{min}}(z,M)}^{r_{\text{max}}(z)}  dr \frac{d^{2}P}{dMdr} \nonumber \\
                                                &\times \frac{dP}{d\log_{10}L}(\langle L_{\text{LW, Ren.}}\rangle, J, r). \nonumber
\end{align}
Here $\frac{d^{2}P}{dMdr}(z, M, r)$ (units: h$^{2}$ M$_{\odot}^{-1}$ Mpc$^{-1}$) is the probability density used to count the number of neighbouring halos in a given mass-separation bin, $\frac{dP}{d\log_{10}L}(\langle L_{\text{LW}}(z, M)\rangle, J, r)$ is the probability density used to assign a LW luminosity density to each halo and we use the Heaviside function $\Theta[r - r_{\text{s}}(z, M, t_{\text{ff}}(z))]$ to account for metal pollution via supernova outflows from a neighbouring halo in the fiducial model. Now we define each term in the integral over $M$ and $r$ below.

\subsubsection{Counting the Neighbouring Halos}
If the neighbouring halos were uniformly distributed in space, then the probability of finding a neighbouring halo with halo mass $[M, M + dM/2]$ and physical separation $[r, r+dr/2]$ from the target halo would be:
\begin{align}
\frac{d^{2}P}{dMdr}(z, M, r) dM dr &= 4\pi r^{2} (1+z)^{3}\\
                                   &\times \frac{dn_{\text{SMT}}}{dM}(z,M)\, dM dr.\nonumber
\end{align}
Here $4\pi r^{2} dr$ is the physical volume of the shell surrounding the target halo, $\frac{dn_{\text{SMT}}}{dM}(z,M) dM$ is the halo number per unit comoving volume and $(1+z)^{3}$ converts from it from comoving to physical volume. But this only holds for a comoving separation $r_{\text{co, max}} \gtrapprox 100 \, \text{cMpc}$.\footnote{We computed this by finding $r_{\text{co, max}}$ such that $\xi_{\text{mm}} < 0$ for $r > r_{\text{co, max}}$ and halos are no longer correlated.} We must additionally account for dark matter clustering leading to deviations from the mean matter density $\Bar{\rho}_{\text{m}}$ at close range. We denote this deviation $\delta(z, r_{\text{co}})$ by:
\begin{equation}
    1 + \delta(z, r_{\text{co}}) = \frac{\rho_{\text{m}}(z, r_{\text{co}})}{\Bar{\rho}_{m}(z)}.
\end{equation}
Rather than computing $\delta(z, r_{\text{co}})$ directly, we find a related quantity: the dimensionless two-point halo-halo correlation function $\xi_{\text{hh}}(z, \text{M}_{\text{min}}(z), M, r_{\text{co}})$. This accounts for the excess probability of finding a neighbouring halo of mass $M$ at a comoving separation of $r_{\text{co}}$ from our target halo of mass M$_{\text{min}}(z)$, both at redshift $z$. We can separate the $M$-dependence using the dimensionless halo bias terms $b(z, M)$ and the $r_{\text{co}}$-dependence using the dimensionless two-point matter-matter correlation function $\xi_{\text{mm}}(z,r_{\text{co}})$ \citep{van_den_Bosch_2013}:
\begin{equation}
\begin{aligned}
    \xi_{\text{hh}}(z, \text{M}_{\text{min}}(z), M, r_{\text{co}}) &\approx b(z, \text{M}_{\text{min}}(z))b(z, M) \\ 
    &\times \xi_{\text{mm}}(z,r_{\text{co}}).
\end{aligned}
\end{equation}
The redshift and mass dependence of the halo bias terms $b(z,M)$ are determined by a fitting function developed by \cite{SMT_2001}. We can later relate physical and comoving separations by:
\begin{equation}
    r = \frac{1}{1+z}r_{\text{co}}.
\end{equation}
The quantity $\xi_{\text{mm}}(z,r_{\text{co}})$ is found by taking the inverse Fourier transform of the nonlinear power spectrum $P(z,k)$ in spherical coordinates where we assume the power spectrum is spherically symmetric:
\begin{equation}\label{eqn:xi_long}
    \xi_{\text{mm}}(z,r_{\text{co}}) = \frac{1}{2\pi^{2}}\int_{0}^{\infty} \text{d}k \, k^{2} P(z,k)\frac{\sin(kr_{\text{co}})}{kr_{\text{co}}}.
\end{equation}
We note that the linear power spectrum $P(z, k)$ varies with $z$ as:
\begin{equation}
    P(z,k) = d(a(z))^{2} \, P(z=10,k), 
\end{equation}
where $a(z) = 1/(1+z)$ and $d(a)$ is the normalised linear growth factor at $z=10$ (\citet{Lukic_2007}):
\begin{equation}
d(a) = \frac{D^{+}(a)}{D^{+}(a=1/(1+10))},
\end{equation}
\begin{equation}
D^{+}(a) = \frac{5\Omega_{m,0}}{2} \, \frac{H(a)}{H_{0}} \, \int_{0}^{a} \frac{da'}{\left[a'H(a')/H_{0}\right]^{3}}.
\end{equation}

\noindent Thus we can approximate the redshift dependence of the matter-matter correlation function as:
\begin{equation}\label{eqn:xi_z_dep}
    \xi_{\text{mm}}(z,r_{\text{co}}) = d(a(z))^{2} \, \xi_{\text{mm}}(z=10,r_{\text{co}}). 
\end{equation}

We create a fitting function based on arrays of $r_{\text{co}}$ and $\xi_{\text{mm}}(z=10,r_{\text{co}})$ values. This allows us to find $\xi_{\text{mm}}(z,r_{\text{co}})$ using Eq. \ref{eqn:xi_z_dep}, rather than computing it via integration as in Eq. \ref{eqn:xi_long} which is much more computationally heavy.

For brevity, we shall refer to $\xi_{\text{hh}}(z, \text{M}_{\text{min}}(z), M, r_{\text{co}})$ as $\xi_{\text{hh}}$. Finally, the probability of finding a neighbouring halo with halo mass $[M, M + dM/2]$ and physical separation $[r, r+dr/2]$ is given by:
\begin{align}
    \frac{d^{2}P}{dMdr}(z, M, r) \, dM dr &= 4\pi r^{2} (1+z)^{3}\\ 
                                       &\times \frac{dn_{\text{SMT}}}{dM}(z, M)[1+\xi_{\text{hh}}] \, dM dr. \nonumber
\end{align}

For the fiducial model, we integrate over the halo mass range $[M_a, M_{b, \text{fid.}}] = [\text{M}_{\text{min}}(z), 10^{15}$ \msolarc]. For the model informed by \texttt{Renaissance} data, we integrate over the range $[M_a, M_{b, \text{Ren.}}] = [\text{M}_{\text{min}}(z), 10^{9}$ \msolarc] since no halos were found above this mass range for the redshifts considered (11.6 $\leq z \leq$ 18.6). We define the minimum physical separation, $r_{\text{min}}(z, M)$, as $r_{\text{min}}(z, M) = 2r_{\text{vir}}(z,M)$ where $r_{\text{vir}}$ is the virial radius of the neighbouring halo (\cite{Johnson_2012}). This avoids a neighbouring halo overlapping with the target halo.
\begin{equation}
\begin{aligned}
    r_{\text{vir}}(z,M) &= (7.84\times 10^{-4})\Big(\frac{M}{10^{8} \, \text{h}^{-1}\text{M}_{\odot}}\Big)^{1/3}\Omega_{\text{m,0}}^{-1/3}\\
    &\times\Big(\frac{1+z}{10}\Big)^{-1} \, \text{Mpc/h}.
\end{aligned}
\end{equation} 
We set $r_{\text{max}}(z) = r_{\text{co, max}}/(1+z)$ as the maximum physical separation. We found $r_{\text{co, max}} \approx 119.448$ cMpc/h, which is the maximum comoving separation where $\xi_{\text{mm}}(z=10.0, r_{\text{co}}) > 0$. Beyond this value, a distant halo would no longer be correlated with the target halo. 

\subsubsection{Assigning Lyman-Werner Luminosity Density} \label{subsubsec:assigning}
The probability density that a neighbouring halo of mass $M$ (units: h$^{-1}$ M$_{\odot}$ and physical separation $r$ (units: Mpc/h) has a LW luminosity density $L$ is given as:
\begin{align}\label{eqn:L_pd}
    \frac{dP}{d\log_{10}L}(\langle L_{\text{LW}}\rangle, J, r) &= \frac{1}{\sigma_{\text{LW}}\sqrt{2\pi}}\\&\times\exp\Bigg[\frac{-(x - \mu)^{2}}{2\sigma_{\text{LW}}^{2}}\Bigg],\nonumber
\end{align}
where $\sigma_{\text{LW}} = 0.4$, $x = \log_{10}L$ ($L$ in units of 10$^{26}$ erg s$^{-1}$ Hz$^{-1}$), $L = 16\pi^{2}r^{2}J$ and $\mu = \log_{10}\langle L_{\text{LW}}(z, M)\rangle$ (adapted from Eq. 3 of \cite{Dijkstra_2008}). A LW luminosity density (and LW flux) is assigned to each halo such that they follow a lognormal distribution in $r$ with a mean LW luminosity density $\langle L_{\text{LW}}(z, M) \rangle$. In our fiducial model, we adapt the model used by D08 and D14.

To compute $\langle L_{\text{LW, fid.}}(z, M) \rangle$, we must first consider the mean LW photon production rate $\langle Q(t) \rangle$ (units: h s$^{-1}$ M$_{\odot}^{-1}$):
\begin{equation}
    \langle Q(t) \rangle = (Q_{0})[1+(t_{6}/4)]^{-3/2}e^{-t_{6}/300},
\end{equation}
where $Q_{0} = (10^{47}/\text{h})$ photons h s$^{-1}$ \msolar$^{-1}$ and $t = (t_{6}) \, (10^{6} \, \text{yr})$. This quantity is derived from the Starburst99 population synthesis model developed by \cite{Leitherer_1999} while assuming that star formation occurs with a Salpeter initial mass function.

The mean LW luminosity density $\langle L_{\text{LW, fid.}}(t, M)\rangle$ (units: erg s$^{-1}$ Hz$^{-1}$) of a halo of mass $M$ at a time $t$ after star formation begins is given as:
\begin{equation}\label{eqn:mean_L}
     \langle L_{\text{LW, fid.}}(t, M) \rangle = \frac{h_{\text{P}}\langle\nu\rangle}{\Delta\nu}\langle Q(t)\rangle f_{\text{esc}}f_{*}\frac{\Omega_{\text{b}, 0}}{\Omega_{\text{m}, 0}}M,
\end{equation}
where $h_{\text{P}}$ is Planck's constant, $\langle \nu \rangle$ is the mean LW frequency, $\Delta\nu$ is the LW frequency range, $f_{\text{esc}} = 1$ is the LW photon escape fraction and $f_{*} = 0.05$ is the star formation efficiency.

\begin{figure}
    \centering
    \includegraphics[scale=0.55]{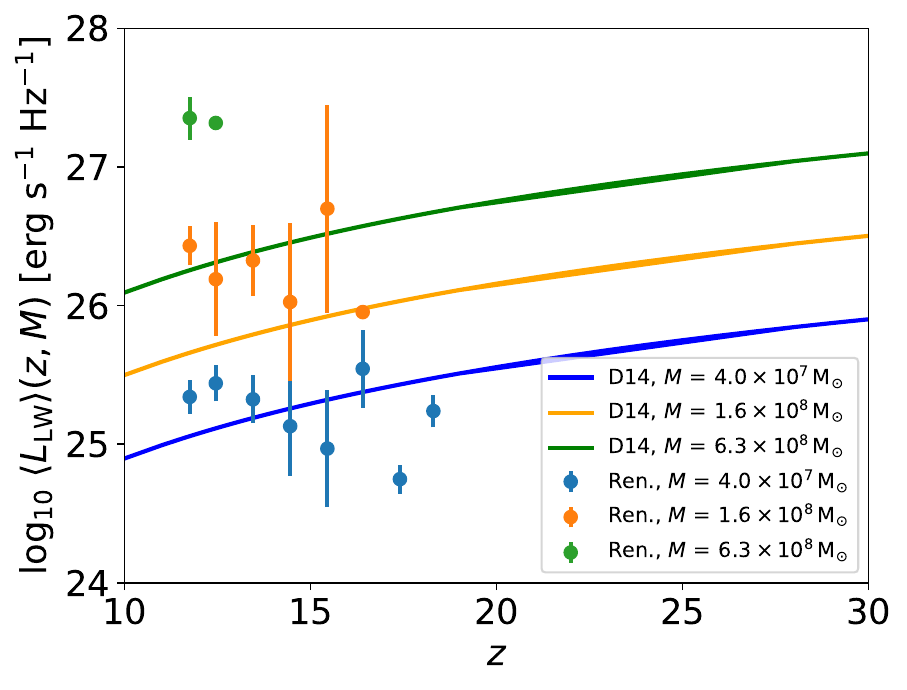}
    \caption{Mean LW luminosity density vs. redshift for a number of halo masses. This is the luminosity density, in units of erg s$^{-1}$ Hz$^{-1}$, emitted by halos with the masses shown in the legend. The mean LW luminosity density in the fiducial model is given by Eq. \ref{eqn:mean_L}. This quantity increases as time decreases and it is evaluated at the free-fall time. The free-fall time decreases as redshift increases, thus the mean luminosity increases as redshift increases. We are evaluating the luminosity when the  stars are younger (and hence more luminous) at higher redshift. We also plot data points from the \texttt{Renaissance} simulation suite with the mean LW luminosity emitted from halos of a given mass as a function of $z$. In this case the mean LW luminosity is almost flat with a small decrease (i.e. lower star formation efficiencies) seen for the lowest mass halos. The \texttt{Renaissance} data points shown are the median values within a redshift bin, with the error bars being the interquartile range.}
    \label{fig:meanL_plot}
\end{figure}

For possible heavy seed formation, it is imperative that the gas within the target halo fully collapses before it may cool and fragment to form stars. It must receive supercritical LW flux from its neighbouring halos for the duration of its free-fall time $t_{\text{ff}}(z)$. The free-fall time is given as: 
\begin{equation}
    t_{\text{ff}}(z) = \sqrt{\frac{3\pi}{32G\rho(z)}} \sim 83\Bigg[\frac{1+z}{11}\Bigg]^{-\frac{3}{2}} \text{Myr},
\end{equation}
where the density of a halo $\rho(z) \approx 200\rho_{\text{m}}(z)$ and we assume that star formation begins in all neighbouring halos simultaneously. Since $\langle L_{\text{LW, fid.}}(t, M) \rangle$ monotonically decreases as $t$ increases, if the target halo receives supercritical LW flux at $t=t_{\text{ff}}(z)$, then it has received supercritical LW flux at $t<t_{\text{ff}}(z)$.

Therefore we evaluate $\langle L_{\text{LW, fid.}}(t, M) \rangle$ at $t = t_{\text{ff}}(z)$, making it strictly a function of $z$ and $M$ only. Figure \ref{fig:meanL_plot} depicts how $\langle L_{\text{LW, fid.}}\rangle$ varies with $z$ for three different values of $M$. The different values of $M$ are marked as solid lines and vary from $4 \times 10^7$ M$_{\odot}$ up to $6 \times 10^8$ \msolarc. What we see is that for the analytic model the mean luminosity increases with redshift - primarily driven by the dependence on the free-fall times which depends on the redshift $z$.

To test the physicality of this model we again appeal to \texttt{Renaissance} and plot the mean LW luminosity $\langle L_{\text{LW, Ren.}}\rangle$ as a function of halo mass and redshift from \texttt{Renaissance}. When calculating the LW flux from \texttt{Renaissance} halos we first determine the stellar mass of that halo and from that calculate the mean LW flux that is produced by that stellar mass according to the \texttt{Renaissance} model. While the data for the largest halo masses is relatively sparse (green data points), the data for the smaller halos is well sampled. In this case we see a relatively flat (slightly decreasing) LW luminosity as a function of redshift. For the lowest mass halos we would expect a slightly lower star formation efficiency and hence a lower mean LW luminosity. However, what we do not see, and is a limitation of the analytic model, is a mean LW luminosity which increases with redshift over this range. This divergence of the analytic and hydrodynamical models will feed into our results. Having now introduced the methodology behind our analysis we now present our results. 

\section{Results}\label{Sec:Results}
As previously stated, the aim of this work is firstly to reproduce and verify the results from D14 and secondly to compare the results of this analysis against the results of other numerical experiments from the literature.  We augment this goal by also taking advantage of the \texttt{Renaissance} suite of simulations and use some of the relevant \texttt{Renaissance} data in the analytic models. Although the \texttt{Renaissance} suite cannot capture the rare halos that experience super-critical LW radiation at the values thought necessary to produce heavy seeds (i.e. $J_{\text{crit}} \gtrsim 300$ J$_{21}$), it is nonetheless an important check on the self-consistency of the analytic model, particularly the probability of finding pristine halos $P_{\text{pristine}}(z)$ as a function of redshift and the mean LW luminosity density $\langle L_{\text{LW}}(z, M) \rangle$ as a function of halo mass and redshift. 

\begin{figure}
    \centering
    \includegraphics[scale=0.5]{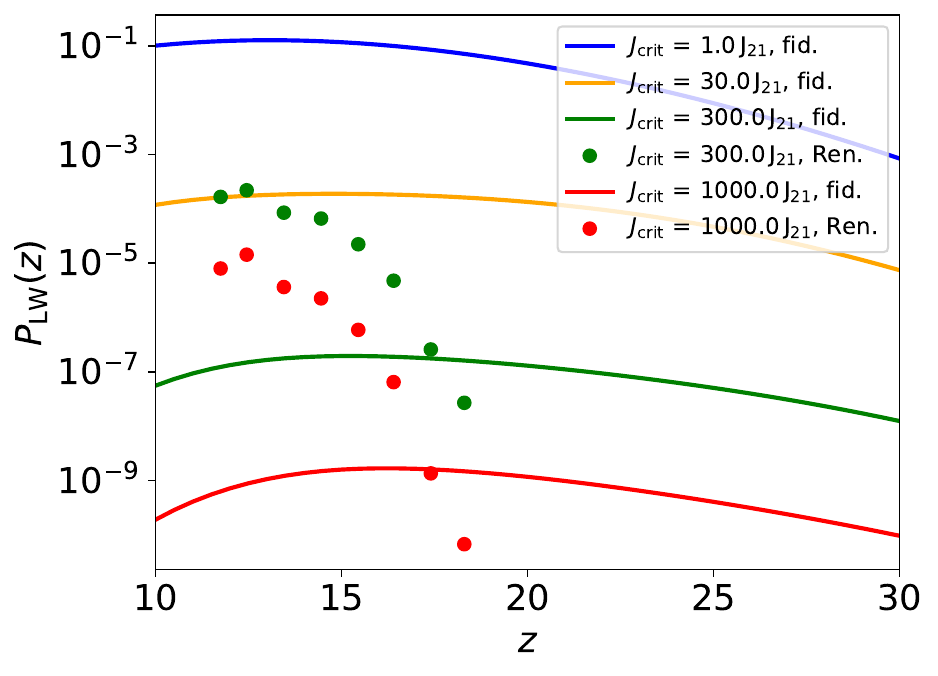}
    \caption{The supercritical probability vs. redshift. The probability of a halo receiving a super-critical LW flux is given by the y-axis. Line colours refer to values of $J_{\text{crit}}$. As expected, the probability of a halo receiving a high flux (e.g. $J_{\text{crit}} \geq 300$ J$_{21}$) is low. The solid lines are from the analytic model. The points are from the \texttt{Renaissance}-informed models. For these models, the probability of a halo receiving a super-critical flux drops sharply with redshift - more in line with expectations.}
    \label{fig:P_Jcrit_plot}
\end{figure}

\begin{figure*}
    \centering
    \includegraphics[scale=0.8]{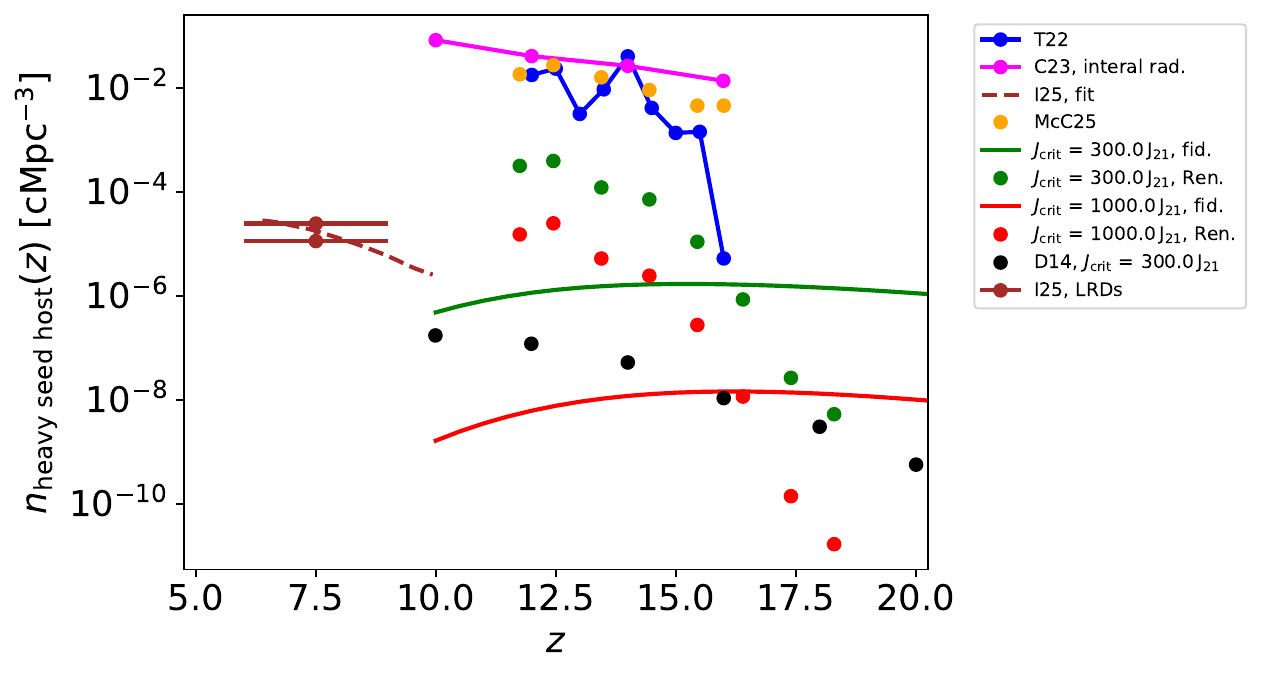}
    \caption{Number density of heavy seeds vs. redshift. Black dots are original points taken from D14, green and red lines show our use of the same analytic model methodology as outlined in D14, the green and red dots use \texttt{Renaissance}-informed data as part of the analytic model, orange dots are from \cite{McCaffrey_2024}, the solid blue line with circular markers is from \cite{Trinca_2022}, the solid magenta line with circular markers is from \cite{Chiaki_2023} (with internal radiation), and the brown lines are from \cite{Inayoshi_2025} (dashed line for the AGN number density fit; dots for the LRD number density estimate based on observations). The LW-only channels models (solid lines, black dots, green and red dots) are unable or only very marginally able to reproduce recent JWST high-$z$ AGN number density of $\gtrsim 10^{-4}$ cMpc$^{-3}$. On the other hand, the models of \cite{McCaffrey_2024} and \cite{Trinca_2022} produce higher number densities and are thus more compatible with recent JWST observations.}
    \label{fig:Number_density_plot}
\end{figure*}

\subsection{Supercritical Probability}
In previous works, $J_{\text{crit}}$ was chosen to be 30 -- 300 $\text{J}_{21}$ for a T = $10^{4} \, \text{K}$ blackbody spectrum (see \cite{Shang_2010}) and 1000 $\text{J}_{21}$ for a T = $10^{5} \, \text{K}$ spectrum (see \cite{WolcottGreen_2011}). D14 chose $J_{\text{crit}} = 300 \, \text{J}_{21}$ as an intermediate value between these two spectra. In this work we follow D14 and choose  $J_{\text{crit}} = 300 \,\text{J}_{21}$ as a critical threshold for the LW pathway. We also investigate solutions with $J_\text{crit} = $ 1.0, 30.0 and 1000.0 $\text{J}_{21}$.

In Figure \ref{fig:P_Jcrit_plot}, we plot the probability of a target halo receiving a supercritical flux at different redshifts. The probability is plotted for a range of different values of the critical flux, $J_{\text{crit}}$, from $J_{\text{crit}}$ = 1.0 J$_{21}$ up to $J_{\text{crit}}$ = 1000.0 J$_{21}$. Solid lines are from the analytic model. As expected the probability of a target halo receiving a flux in excess of $J$ = 1.0 J$_{21}$ is very high. In this case by a redshift of $z = 10$, approximately 1 in 10 halos meeting the target halo criteria will receive a flux greater than or equal to $J$ = 1.0 J$_{21}$. However, the probability of a target halo receiving significantly higher fluxes is much less. For example the probability of a target halo receiving a flux in excess of $J$ = 1000.0 J$_{21}$ at $z = 10$ is less than 1 in $10^9$. Note that in this model, the peak in terms of probabilities occurs at $z$ $\sim$ 15 and decreases at higher and lower redshifts.

It should also be noted that the probability of a halo receiving a supercritical flux does not decrease as rapidly as perhaps expected towards very high redshift (i.e. $z$ $\ge 20$). This is an inherent characteristic of the fiducial model. The model requires that a target halo is illuminated by a super-critical flux for a free-fall time (see \S \ref{subsubsec:assigning}). While the number of star-forming halos decreases as per the halo mass function, the free-fall time shrinks dramatically with redshift. The two effects cancel each other out somewhat and hence the probability of receiving a super-critical flux does not decrease as redshift increases as rapidly as expected. 

Overplotted in Figure \ref{fig:P_Jcrit_plot} are the results when applying information from the \texttt{Renaissance} datasets (median values of each redshift bin). In this case the mean LW values $\langle L_{\text{LW, Ren.}}(z, M) \rangle$ are used instead of the analytic model values $\langle L_{\text{LW, fid.}}(z, M) \rangle$ and we exclude the Heaviside function from the first line of Eq. \ref{eqn:dP_dlog10J}, i.e. we do not account for the influence of metal pollution in this plot but we do in Figure \ref{fig:Number_density_plot}. For the simulation values, we see a much steeper decline in the probability of a halo receiving a super-critical flux (driven mainly by the mean LW luminosity density). Hence, for redshifts approaching $z \sim 20$ the probability of a halo receiving a supercritical flux is negligible. This is in comparison to the analytic model where the probabilities only decline slowly. 

\subsection{Heavy Seed Number Density}
Putting everything together, we now plot the number density of halos acting as heavy seed formation sites as a function of redshift in Figure \ref{fig:Number_density_plot}. In total we plot 10 datasets. With the exception of the black dots, all dotted datasets are the median values of a redshift bin. These datasets include: our fiducial model with $J_{\text{crit}} = 300 \, \text{J}_{21}$ (green line) and $J_{\text{crit}} = 1000 \, \text{J}_{21}$ (red line); our \texttt{Renaissance}-augmented model with $J_{\text{crit}} = 300 \, \text{J}_{21}$ (green dots) and $J_{\text{crit}} = 1000 \, \text{J}_{21}$ (red dots); the heavy seed number density from \cite{Trinca_2022} (solid blue line with circular markers); the DCBH number density from \cite{Chiaki_2023} (solid magenta line with circular dots); the heavy seed formation site number density from \cite{McCaffrey_2024} (orange dots); and the heavy seed formation site number density from D14 with $J_{\text{crit}} = 300 \, \text{J}_{21}$ (black dots).

We also include an AGN number density estimate from \cite{Inayoshi_2025}. This model is based on the observed and inferred abundance of LRDs, suggesting that LRDs are a phase in galaxy evolution. The brown dashed line in Figure \ref{fig:Number_density_plot} shows this fit for $6\leq z \leq 10$. The brown dots are the inferred AGN number densities based on recent observations for $6.5 \leq z \leq 8.5$ from \cite{Kokorev_2024} and \cite{Kocevski_2024}. The fit from \cite{Inayoshi_2025} was normalised such that the AGN number density matches the observed LRD number density of $\approx 3 \times 10^{-5}$ cMpc$^{-3}$ for $z \sim 4 - 7$. We include these estimates to provide context for our heavy seed datasets but we must note that a heavy seed number density cannot be directly compared to these AGN number density estimates. These AGN number densities which are informed by recent observations likely account for only the most active of AGN, while in reality many AGN will be quiescent and thus making the total AGN number density larger. We also note that only a fraction of heavy seeds will undergo efficient growth to reach SMBH masses by $z \sim 6$. Thus in Figure \ref{fig:Number_density_plot}, if a heavy seed number density estimate at $z \sim 10$ is equal to or greater than this AGN number density estimate, that pathway may be viable to account for SMBH number densities. If a heavy seed number density estimate is less than the AGN number density estimate, that pathway can at best only account for a fraction of SMBHs.

For our fiducial models (green and red lines), we are unable to match results from D14 (black dots) exactly despite following their methodology and our results deviate from theirs, particularly at high redshift. Without access to their code base we cannot determine where the discrepancy arises. Our codebase and pipeline will be publicly available on GitHub but until then may be accessed upon request. 

Both the results from our fiducial model and the results of D14 do however agree that the number density of heavy seed black holes (or indeed heavy seed hosting halos) is less that $10^{-6}$ cMpc$^{-3}$ at $z \gtrsim 10$. With the number density of MBH hosting galaxies at least two orders of magnitude greater than this (and potentially likely much higher) at $z \gtrsim 4$ \citep[e.g.][]{Perez-Gonzalez_2024, Inayoshi_2025}, the LW channel is likely unable to explain the high abundances of MBHs in the early Universe based on our fiducial model. This is the first takeaway from our analysis.

We plot the results using values obtained from the \texttt{Renaissance}-informed analytic models as green ($J_{\text{crit}} = 300 \, \text{J}_{21}$) and red dots ($J_{\text{crit}} = 1000 \, \text{J}_{21}$). In this case the results are more encouraging - albeit with a steeper decline. Note that we are not using the \texttt{Renaissance} output to make predictions but use their mean LW luminosity density values $\langle L_{\text{LW, Ren.}}(z, M) \rangle$ and the impact of genetic and external metal pollution is informed by their $P_{\text{pristine, Ren.}}(z)$ function (see Figure \ref{fig:meanL_plot} and Figure \ref{fig:P_pristine}). We see that the number density of heavy seed hosting halos peaks at approximately $10^{-4}$ cMpc$^{-3}$ at $z \sim 12.5$ for $J_{\text{crit}}$ = 300 J$_{21}$. These numbers are on the face of it consistent, albeit marginally, with the recent results from JWST (see AGN number density estimates from \cite{Inayoshi_2025} (brown dots and brown dashed line). However, our models are for the seeds and not the candidate AGN detected by JWST. Given the growth requirements of the seeds combined with the expected duty cycle of AGN, these number densities are still likely incompatible with current JWST observations. 

We also plot more recent results by \cite{Trinca_2022} (solid blue line with circular markers), \cite{Chiaki_2023} (solid magenta line with circular dots) and \cite{McCaffrey_2024} (orange dots). \cite{McCaffrey_2024} used the \texttt{Renaissance} simulations to analyse the formation and later the growth of heavy seed black holes while accounting for rapid assembly and merger history of halos. We do not account for growth in this comparison and only consider their number density of heavy seed formation sites. They used the following criteria when identifying halos that may host massive black holes:
\begin{itemize}
    \item $\text{T}_{\text{vir}} \gtrsim 10^4$ K, 
    \item $Z < 10^{-3} \, \text{Z}_{\odot}$,
    \item $M(r < 0.5R_{\text{vir}})/M(r < R_{\text{vir}}) > 0.5$,
    \item $\dot{M} > 0.1 \, \text{M}_{\odot} \, \text{yr}^{-1}$.
\end{itemize}
The results from \cite{McCaffrey_2024} show that the formation of heavy seed MBHs via the so-called rapid assembly channel results in significantly higher number densities with values of approximately $10^{-2}$ cMpc$^{-3}$ or higher. These results from \cite{McCaffrey_2024} are consistent with recent JWST results (again with the caveat that the growth of seeds and the duty cycle of AGN will push these number densities downwards). 

\cite{Trinca_2022} performed a similar analysis to \cite{McCaffrey_2024} using the \textsc{Cosmic Archaeology Tool} (\textsc{CAT}), a semi-analytic model which allowed them to follow the formation of the first stars and black holes while accounting for accretion and mergers. They also account for how star formation in mini-halos can be affected by LW flux from nearby highly star-forming galaxies. This inhibits H$_{2}$ cooling and potentially sterilises these halos until they reach the atomic-cooling regime. In their model, if a dark matter halo meets the following conditions:
\begin{enumerate}
    \item $\text{T}_{\text{vir}} \gtrsim 10^4$ K, 
    \item $Z < 10^{-3.8} \, \text{Z}_{\odot}$, 
    \item $J_{\text{LW}}>300 \, \text{J}_{21}$,
\end{enumerate}
then a heavy seed black hole with $M_{\text{BH}}=10^5$ M$_{\odot}$ is set at the centre of the galaxy within the halo. Their data in Figure \ref{fig:Number_density_plot} indicates the number density of newly-formed heavy seeds at a given redshift. Their results show similar number densities to \cite{McCaffrey_2024} with peak values $~ 10^{-2}$ cMpc$^{-3}$. Their model shows number densities significantly beyond the LW-only channel, again compatible with more recent results from JWST. 

When computing the number density of DCBHs, \cite{Chiaki_2023} consider both external and internal LW radiation, rather than only external sources like we do. They make a direct comparison to \cite{Trinca_2022} by considering halos with $\text{T}_{\text{vir}} > 10^4$ K, $J_{\text{crit}}=300$ J$_{21}$ and $Z < 10^{-3.8}$ Z$_{\odot}$. Internal radiation from a single halo depends on its LW photon emissivity (from both Pop III and Pop II stars) and the radius of the galaxy hosted by that halo (where the galaxy radius is 10\% of the halo virial radius). External radiation received by a single halo is computed from the total LW flux from all other halos and is dependent on LW photon emissivity and the distance between the target and source halos. Metal pollution is accounted for via chemical evolution equations of Pop I/II stars and SN winds. Their highest number density value is almost $10^{-1}$ cMpc$^{-3}$ at $z=10$. 

Additionally, \cite{Bhowmick_2021} explored a similar LW channel with the inclusion of low gas angular momentum to track SMBH seed formation and \cite{Bhowmick_2024} used the \texttt{BRAHMA} simulations to track DCBH formation. Both works found that more efficient heavy seeding channels may be necessary to account for the most massive black holes at high redshift.

In summary, Figure \ref{fig:Number_density_plot} tells us that both D14 and our re-implementation of the D14 model show results which are incompatible with recent JWST data. Other pathways investigated by \cite{McCaffrey_2024},\cite{Trinca_2022} and \cite{Chiaki_2023} appear more promising to explain the overall MBH population. A similar result was shown by \cite{Bhowmick_2021}.

\section{Discussion and Conclusions}\label{Sec:Discussion}
In this paper, we have reviewed the analytic model of \cite{Dijkstra_2014} in terms of calculating the number density of MBHs that can be formed through the so-called Lyman-Werner (LW) channel. In this framework, a super-critical flux of LW irradiates a target halo. The target halo must have a mass exceeding the atomic-cooling threshold and must be metal-free. The super-critical flux required can vary from halo to halo but is likely to be excess of 300 J$_{21}$. Such a high value can only be produced by a nearby neighbouring halo. A weakness of this model is assuming that there is no correlation between a halo being metal-free and receiving a super-critical LW flux i.e. we multiply the probabilities $P_{\text{pristine}}(z)$ and $P_{\text{LW}}(z, J_{\text{crit}})$. More realistically, a super-critical LW flux may suppress star formation in a progenitor halo and thus the target halo has an increased probability of being metal-free. Such an analysis of the merger history of the target halos is beyond the scope of this study. In agreement with D14, we find that the number density of target halos receiving a  critical flux in excess of 300 J$_{21}$ is approximately $10^{-6}$ cMpc$^{-3}$ at $z = 10$ (see Figure \ref{fig:Number_density_plot}). The number densities drop, as expected, towards higher redshifts. 

Despite considerable effort, we were unable to reproduce the exact results of D14 and our fiducial model results differ from the D14 results at $z \gtrsim 20$. We varied the cosmology implemented, the integration limits for neighbouring halo mass $M$ and separation $r$, how the halo-halo correlation was computed and the definition of the mean LW luminosity density. Our analysis tools and pipeline will be available on GitHub. Our analysis, following the methodology of D14, shows that the number density of heavy seeds is almost constant out to very high redshift ($z \gtrsim$ 20) - this is primarily due to how the mean LW luminosity $\langle L_{\text{LW, fid.}}(z, M) \rangle$ produced by a halo is calculated in the D14 model. Nonetheless, our analysis agrees very well with the D14 as we approach $z = 10$. However, in both cases the number densities remain close to or below $10^{-6}$ cMpc$^{-3}$. 

To check the physical consistency of the D14 model, we augment the analytic model with information taken directly from the \texttt{Renaissance} simulation suite. Specifically, we take data of the mean LW luminosity density $\langle L_{\text{LW, Ren.}}(z, M) \rangle$ produced as a function of redshift and halo mass and the pristine fraction $P_{\text{pristine, Ren.}}(z)$ of halos as a function of redshift (see Figures \ref{fig:meanL_plot} and \ref{fig:P_pristine}). Using this augmented model, we find that number density of heavy seed hosting halos increases steeply between $z \sim 18$ and $z \sim 10$. The peak number density of heavy seed hosting halos reaching values close to $10^{-4}$ cMpc$^{-3}$ at $z \sim 10$. The \texttt{Renaissance}-informed models are nonetheless still likely incompatible with the recent results on AGN fractions at high-redshift \citep[e.g.][]{Perez-Gonzalez_2024, Greene_2024, Inayoshi_2025} given subsequent growth requirements of the seeds combined with the expected duty cycle of AGN. \\
 
\indent A slight weakness of this model is the lack of incorporation of baryonic matter streaming velocities \citep{Tseliakhovich_2010}. This would suppress star formation in halos with masses $\lesssim$ 10$^{6}$ M$_{\odot}$ \citep{Tseliakhovich_2011, OLeary_2012, Xu_2014}. This is an order of magnitude below the M$_{\text{min}}$ values we consider, so this may not affect our results greatly. It could push $P_{\text{pristine, Ren.}}(z)$ to higher values since star formation would be suppressed in more progenitor halos, leading to more target halos avoiding genetic metal pollution. 

We have shown through our heavy seed number density predictions that the LW-only channel is likely sub-dominant compared to other channels when accounting for AGN number density estimates based on recent observations. While these observations are still hotly debated and the exact make-up of the JWST galaxies unclear, even if some fraction of the galaxies host AGN (as is strongly suspected) then the LW-only channel cannot be responsible - the predicted number densities are simply too low.

\section*{Acknowledgments}
\noindent HOB thanks past and present members of JR's research group for many helpful discussions and insights: Stefan Arridge, Lewis R. Prole, Saoirse Ward, Pelle van de Bor and Daxal Mehta. JR acknowledges support from the Royal Society and Research Ireland through the University Research Fellow programme under grant number URF$\backslash$R1$\backslash$191132. JR also acknowledges support from the Research Ireland Laureate programme under grant number IRCLA/2022/1165. JHW acknowledges support from NSF grant AST-2108020 and NASA grant 80NSSC21K1053. EV is supported by NSF grant AST-2009309, NASA ATP grant 80NSSC22K0629, and STScI grant JWST-AR-05238. We also thank the anonymous referee for a constructive and insightful report.

\bibliographystyle{mn2e}
\bibliography{references.bib}
\end{document}